\documentclass[aps,prx,groupedaddress, superscriptaddress,longbibliography,reprint]{revtex4-2}

\usepackage{amsmath}  
\usepackage{amssymb}
\usepackage{amsthm}
\usepackage{bbm}
\usepackage{graphicx}
\usepackage{color}
\usepackage{upgreek}
\usepackage[linkcolor = blue, citecolor = blue, urlcolor = blue, colorlinks = true]{hyperref}

\usepackage{amssymb}
\usepackage{bm}
\usepackage[capitalise]{cleveref}
\usepackage{textcomp}
\usepackage{hyperref}
\usepackage[usenames,dvipsnames]{xcolor}
\usepackage[normalem]{ulem}
\usepackage{wasysym}

\usepackage{floatrow}
\usepackage{wrapfig}
\usepackage{pgfgantt}

\usepackage{mathptmx}
\usepackage{bm}

\begin{document}

\title{Fluidic hysterons and memory in flow networks}
\author{Abhineet Singh Rajput}
\author{Amir A. Pahlavan}
\email[]{amir.pahlavan@yale.edu}
\affiliation{Mechanical Engineering and Materials Science, Yale University, New Haven, CT, 06511}

\begin{abstract}
Hysterons provide a minimal description of memory in driven matter: bistable elements with distinct switching thresholds whose interactions generate hysteresis, avalanches, and return-point memory or its violation. Experimental realizations have so far been dominated by solid-state mechanical systems, where bistability is usually encoded structurally through buckling, snap-through, or geometric incompatibility. Here we realize hysteron physics through a hydrodynamic route. A single elastic fiber anchored in a microfluidic channel becomes bistable through nonlinear elastohydrodynamic feedback: viscous loading deforms the fiber, deformation reshapes hydraulic resistance, and flow redistribution modifies the loading. This feedback produces a fluidic hysteron whose onset is organized by a cusp catastrophe in geometric control parameters. A parallel bypass channel acts as a geometric load line that reshapes, and can even eliminate, bistability while simultaneously mediating long-ranged hydraulic interactions between fibers. In arrays, varying a single geometric parameter drives a transition from a non-interacting Preisach regime with return-point memory to an interacting regime with avalanche-like switching and return-point-memory violation. These results establish a passive hydrodynamic route to hysteron networks, in which memory emerges from flow–structure feedback and global hydraulic constraints rather than solid-state multistability or external control.
\end{abstract}

\maketitle

\section{Introduction}

\noindent Living flow networks, from slime molds and fungal mycelia to the microvasculature of the brain, record aspects of their interaction history in their architecture. Flows of nutrients, ions, metabolites, and signaling molecules both transmit information through these networks and remodel them: vessel diameters and mechanical properties adapt in response to shear, pressure gradients, and chemical cues \cite{Tero10,Pries10,Ronellenfitsch16,Roper19,Kramar21,Oyarte25}. This interplay between local adaptation and global hydraulic coupling raises a fundamental question: how do distributed memories emerge in flow networks?

Memory, however, is not a privilege of living systems. Under quasistatic cyclic driving, granular packs, crumpled sheets, amorphous solids, and mechanical metamaterials \cite{Keim14,Galloway22,shohat2022memory,Sirote24,Paulsen25} can be trained so that their response depends on loading history rather than instantaneous load. Their deformation is often smooth until interrupted by abrupt, localized rearrangements, such as yield events in disordered solids or snap-through instabilities in architected structures, that reorganize the internal state \citep{Mungan26}. Each such rearranging region can be idealized as a bistable switch with distinct up and down thresholds: a \textit{hysteron} \cite{preisach1935magnetische,keim2019memory}. When many hysterons act independently, their superposition produces hysteresis and return-point memory (RPM), whereby returning the drive to a previous extremum restores the corresponding prior state \cite{preisach1935magnetische,Sethna93}. When hysterons interact, a flip in one region shifts the thresholds of others, opening new state-transition pathways and enabling cascades of avalanche-like events \cite{van2021profusion,muhaxheri2024bifurcations,shohat2025geometric}. This viewpoint has elevated hysteresis from a material response to a design principle: by arranging and coupling hysterons, one can encode counting, memory storage, and reproducible switching sequences as forms of mechanical information processing \cite{Pashine19,Chen21,Bense21,Kwakernaak23,liu2024controlled,Paulsen26}. Yet nearly all experimental realizations remain solid-state systems, where switching is rooted in snap-through, buckling, or geometric incompatibility. 

Fluidic networks offer a distinct setting for such questions. Kirchhoff-like hydraulic constraints couple distant elements, so that a local change in conductance redistributes pressure and flux throughout the network. At the microscale, however, viscous flows obey the linear Stokes equations and are kinematically reversible \cite{Purcell77}. History dependence therefore requires additional state variables that do not simply rewind under reversal, for example viscoelastic stresses, fluid--fluid interfaces, activity, or deformable boundaries \cite{Groisman03,Prakash07,Fuerstman07,alvarado2017nonlinear,gomez2017passive,Christensen20,Louf20,Park21,peretz2020underactuated,Christov21,garg2024passive,Jorge24}. Existing microfluidic memory and control strategies have exploited such ingredients, often together with external actuation, membrane valves, or pneumatic control \cite{Rhee09,Weaver10,Mosadegh10,Duncan13,ahrar2023pneumatic}. What has remained missing is a passive fluidic counterpart of the hysteron paradigm: an element whose bistability arises intrinsically from quasistatic flow--structure feedback, and whose interactions emerge naturally from global hydraulic constraints.

Here we show that nonlinear elastohydrodynamic feedback between viscous flow and fiber deformation can produce bistability and hysteresis without requiring a structurally bistable element. The minimal realization is a single elastic fiber anchored inside a microfluidic channel: as the imposed flow is ramped quasistatically, the fiber bends and narrows a side gap, redistributing flux between that gap and thin lubrication layers. This closes a feedback loop in which deformation alters the hydraulic resistance, which in turn redirects flow and modifies the hydrodynamic loading. We then introduce a parallel bypass channel as a geometric load-line control that reshapes, and can suppress, bistability while simultaneously tuning long-range hydraulic interactions in fiber arrays. By varying this single geometric parameter, we traverse from a non-interacting Preisach regime to an interacting regime with avalanche-like transitions that violate return-point memory. The same control parameter therefore reshapes the constitutive response of an individual element and the collective interaction topology of the network.

\begin{figure*}[tbh!]
 \centering
\includegraphics[width=1\linewidth]{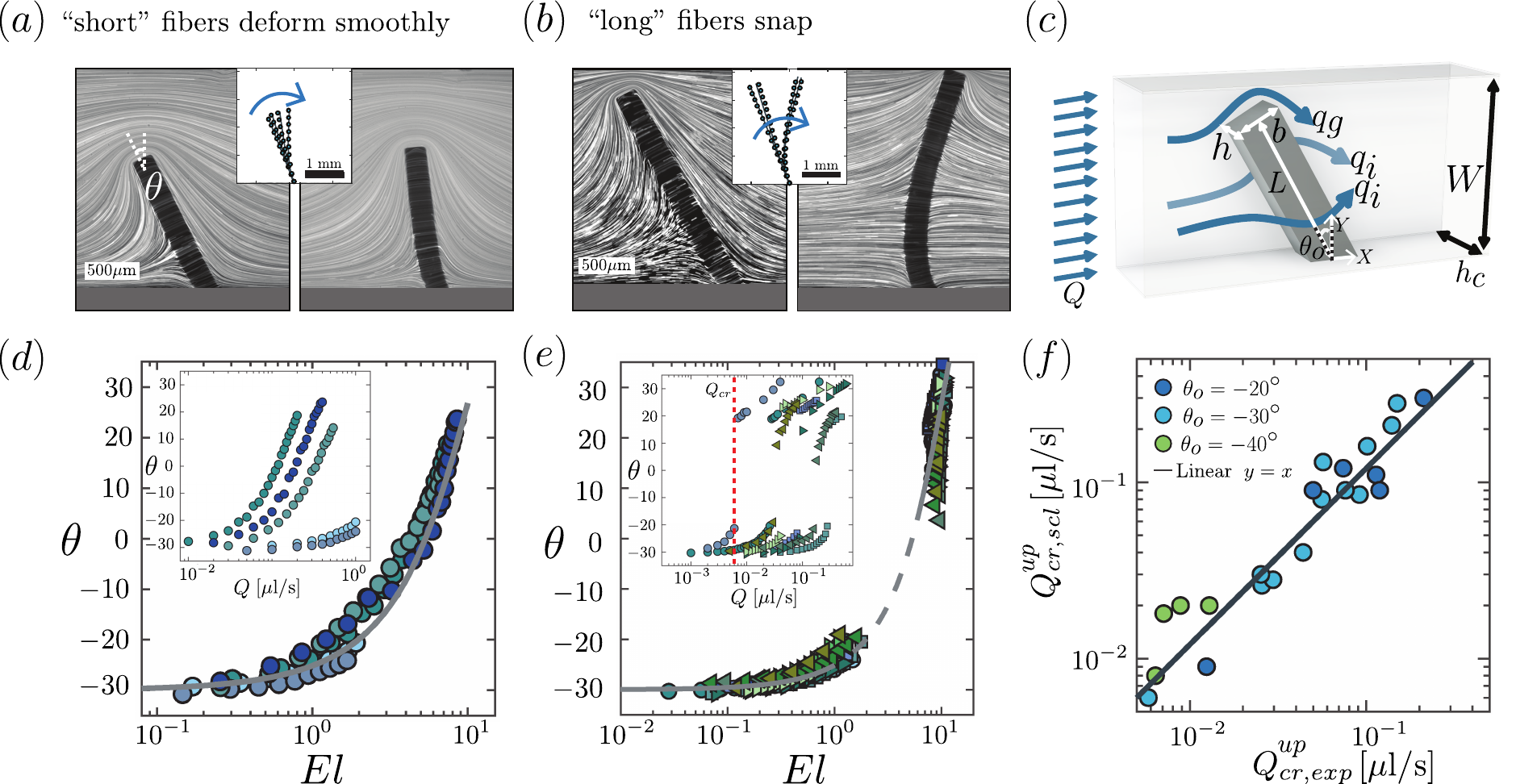}
\caption{\textbf{Elastohydrodynamic snap-through of an anchored fiber.} (a)~Short fiber ($\lambda=L/W=0.65$, $b=200~\mu\mathrm{m}$) in a channel of width $W=2000~\mu\mathrm{m}$ deforms continuously under increasing flow, as characterized by the tip angle $\theta$. (b)~Long fiber ($\lambda=1$, $b=200~\mu\mathrm{m}$) in the same channel undergoes a discontinuous snap. Solid lines show the prediction of Eq.~(1). Insets compare the measured and predicted neutral-axis profiles. (c)~Schematic of the device. A fiber of length $L$ and height $h$ is anchored at angle $\theta_0$ in a channel of width $W$ and height $h_c$. The imposed flow rate $Q$ partitions between the tip gap (flow $q_g$, gap height $h_g$) and the thin lubrication layers (flow $q_i$, layer height $h_i$). (d,e)~Experimental data collapsed using the elastoviscous number $\mathrm{El}$. Short fibers trace a continuous master curve~(d), whereas long fibers split into unsnapped ($\mathrm{El}\lesssim1$) and snapped ($\mathrm{El}\gtrsim10$) branches~(e), with the onset of snapping at $\mathrm{El}\approx1$. Insets show the corresponding dimensional data. (f)~Measured snap-up thresholds $Q_{cr,exp}^{up}$ versus the scaling estimate $Q_{cr,scl}^{up}$ for different anchor angles $\theta_0$ and fibers with $\lambda\simeq1$.}
\label{fig:fig-1} 
\end{figure*}

\section{Elastohydrodynamic bistability}

We pattern a single elastic fiber of length $L$, width $b$, and height $h$ inside a rigid microfluidic channel of width $W$ and height $h_c$, anchored at an upstream tilt angle $\theta_0$ against the flow direction (Fig.~\ref{fig:fig-1} (c)). As the imposed flow rate $Q$ is ramped quasistatically, we image the fiber deformation and track the tip angle $\theta$. Short fibers ($\lambda=L/W=0.65$) deform continuously, whereas long fibers ($\lambda=1$) undergo a discontinuous snap above a critical flow rate $Q_{cr}^{up}$ (Fig.~\ref{fig:fig-1} (a,b) and Supplementary Information Section~1). Varying the fiber length $L$ and width $b$ at fixed $\lambda$ shifts the deformation curves but preserves the distinction between continuous and discontinuous response, showing that the transition is governed by the geometric ratio $\lambda$ rather than by absolute size or stiffness.

This distinction arises from the coupling between flow and deformation. Viscous loading bends the fiber and progressively narrows the lateral gap between the fiber and the adjacent sidewall, increasing the hydraulic resistance of the primary gap pathway, $q_g$. At the same time, fluid passes through thin lubrication layers along the sides of the fiber, whose conductance is approximately constant (Fig.~\ref{fig:fig-1} (c)). As the fiber deflects, flow repartitions between these two pathways, changing the pressure drop and traction acting on the fiber. We capture this feedback with a reduced elastohydrodynamic model that couples lubrication-based hydraulic resistances for the two flow paths to an inextensible Euler--Bernoulli beam description of the fiber (Methods and SI Section 2). Although the fiber itself is not intrinsically bistable, the hydraulic loading depends on its instantaneous shape through the resistance of the narrowing gap, and this nonlinear flow–structure feedback is sufficient to generate bistability. This mechanism differs from conventional fluid-driven snap-through devices, where bistability is already embedded in the elastic structure and the flow merely selects between pre-existing states \cite{gomez2017passive,Ledda25}.

\begin{figure*}[tbh!]
 \centering
\includegraphics[width=1\linewidth]{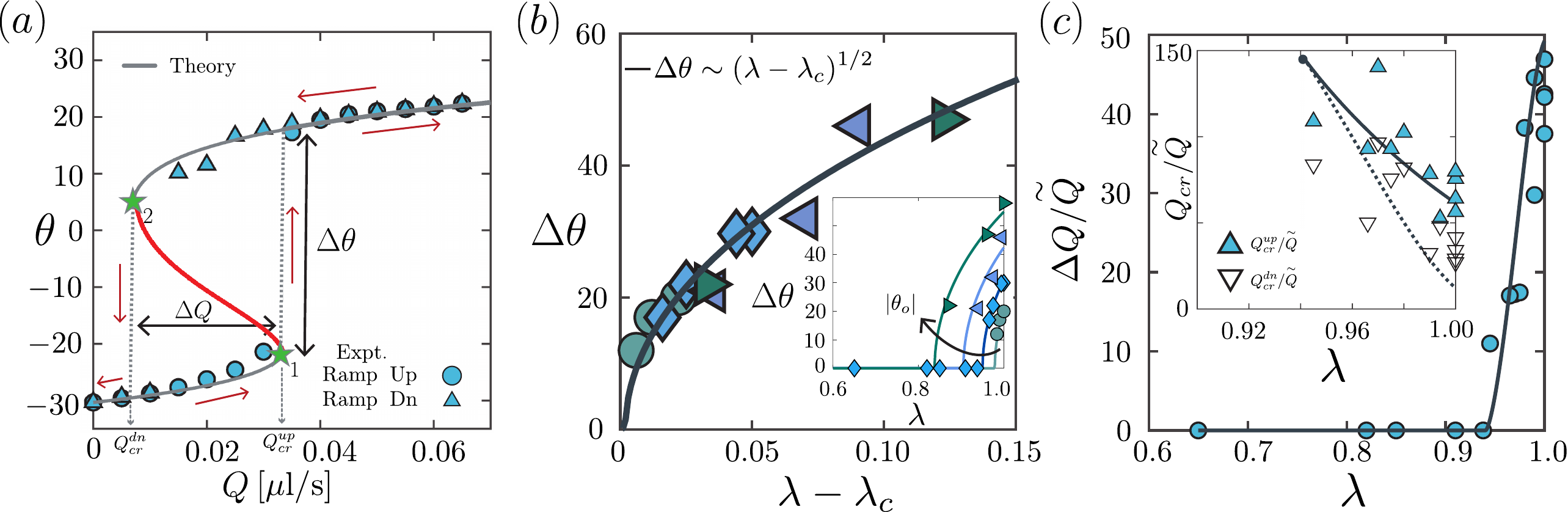}
\caption{\textbf{Constitutive bistability and cusp scaling of a fluidic hysteron.} (a)~Full constitutive response of a snapping fiber with $\lambda=1$ and $b=200~\mu\mathrm{m}$. Under flow-controlled driving, the fiber follows the lower stable branch until the snap-up threshold $Q_{cr}^{up}$, then jumps to the upper branch; on ramp-down it remains snapped until the lower threshold $Q_{cr}^{dn}$, producing hysteresis. The jump amplitude $\Delta\theta$ and hysteresis width $\Delta Q=Q_{cr}^{up}-Q_{cr}^{dn}$ quantify the bistable response. (b)~Jump amplitude $\Delta\theta$ as a function of distance from threshold. Data for different $\theta_0$ collapse when plotted against $\lambda-\lambda_c$, following the square-root scaling predicted by the cusp catastrophe analysis. Inset: $\Delta\theta$ vanishes for $\lambda<\lambda_c$ and becomes finite for $\lambda>\lambda_c$. (c)~Dimensionless hysteresis width $\Delta Q/\tilde{Q}$ (Methods) as a function of $\lambda$. The hysteresis width vanishes below threshold and becomes finite in the snapping regime. Inset: the critical flow rates $Q_{cr}^{up}$ and $Q_{cr}^{dn}$ form the two fold lines that emerge beyond the cusp point. Lines are theoretical model predictions while symbols represent the experimental data.}
\label{fig:fig-2} 
\end{figure*}

Across a wide range of fiber lengths and widths, the measured tip angle $\theta$ collapses onto a single master curve when plotted against the elastoviscous number $\mathrm{El}=\Lambda\mu U_m L^3/D$, where $U_m$ is the mean velocity in the lubrication layers that sets the viscous loading, $\mu$ is the fluid viscosity, and $D=EI$ is the flexural rigidity, with $I=hb^3/12$ the second moment of area and $E$ the fiber modulus. Short fibers trace this curve continuously (Fig.~\ref{fig:fig-1} (d)), whereas long fibers appear to skip an intermediate range of $\mathrm{El}$ (Fig.~\ref{fig:fig-1} (e)): the data cluster near $\mathrm{El}\lesssim1$ in the unsnapped state and $\mathrm{El}\gtrsim10$ in the snapped state. This apparent gap does not reflect a discontinuity in the imposed flow rate $Q$; rather, $U_m$ jumps at snapping as the flow repartitions between the lubrication layers and the narrowing gap, and $\mathrm{El}\propto U_m$ jumps with it. The collapse therefore identifies $\mathrm{El}\approx1$ as the onset criterion for snapping in long fibers. Using this criterion, and assuming that the fiber remains close to $\theta_0$ immediately before snap-up, we infer $U_m$ at onset and map it to a predicted critical flow rate $Q_{cr}^{up}$ (SI Section 3.2). The resulting estimate agrees with the measured snap-up thresholds across different fiber dimensions and anchor angles $\theta_0$ for $\lambda\simeq1$ (Fig.~\ref{fig:fig-1} (f)).

The distinction between continuous and discontinuous response follows from the constitutive structure of the elastohydrodynamic coupling. The steady fiber angle obeys the reduced relation
\begin{equation}
     \tan\theta=\tan\theta_0
     +\frac{\mathrm{El}}{6}\tilde{l}_p^{\,3},
\end{equation}
where $\tilde{l}_p=l_p/L$ is the dimensionless projected fiber length (Supplementary Information Section~2). Under flow-controlled driving, solving for the steady response $\theta(Q)$ yields, for long fibers, a discontinuous jump: as $Q$ increases, the lower stable branch terminates at a critical flow rate $Q_{cr}^{up}$, forcing the system to snap to the upper branch (Fig.~\ref{fig:fig-2} (a)). To reveal the full constitutive relation, we invert the problem and compute the flow required to sustain a prescribed configuration, $Q(\theta)$. This uncovers an \textit{S}-shaped response: two stable branches connected by an intermediate segment with negative slope, $dQ/d\theta<0$. This negative-differential branch is unstable under flow-controlled driving and therefore inaccessible during quasistatic ramping; the fiber instead jumps at the two fold points, which define the snap-up and snap-down thresholds, $Q_{cr}^{up}$ and $Q_{cr}^{dn}$, and hence the hysteresis window $Q_{cr}^{dn}<Q<Q_{cr}^{up}$. This multivalued constitutive law, with two stable branches separated by an unstable negative-slope segment, realizes the constitutive structure of a \textit{hysteron} under flow-controlled driving \cite{muhaxheri2024bifurcations,muhaxheri2025catastrophic}.

The jump amplitude $\Delta\theta$ becomes non-zero only when $\lambda>\lambda_c$, following the scaling $\Delta\theta\sim(\lambda-\lambda_c)^{1/2}$ (Fig.~\ref{fig:fig-2} (b)). This square-root onset is the generic scaling of a cusp catastrophe \citep{Poston78,Strogatz94}, in which two fold lines in the $(\lambda,Q)$ control plane meet at a cusp point $(\lambda_c,Q_{cr})$ (Fig.~\ref{fig:fig-2} (c), inset). The hysteresis width, $\Delta Q\equiv Q_{cr}^{up}-Q_{cr}^{dn}$, likewise becomes finite only beyond $\lambda_c$ and grows with increasing $\lambda$ (Fig.~\ref{fig:fig-2}c). The critical length ratio $\lambda_c$ is set by the anchoring angle $\theta_0$ and is essentially independent of the flexural rigidity $D=EI$ over the explored range. Whether a fiber snaps is therefore a geometric question, determined by $(\lambda,\theta_0)$, whereas the size of the hysteresis window can be tuned by both geometry and stiffness (Supplementary Information Section~4).

\begin{figure*}[tbh!]
 \centering
\includegraphics[width=1.0\linewidth]{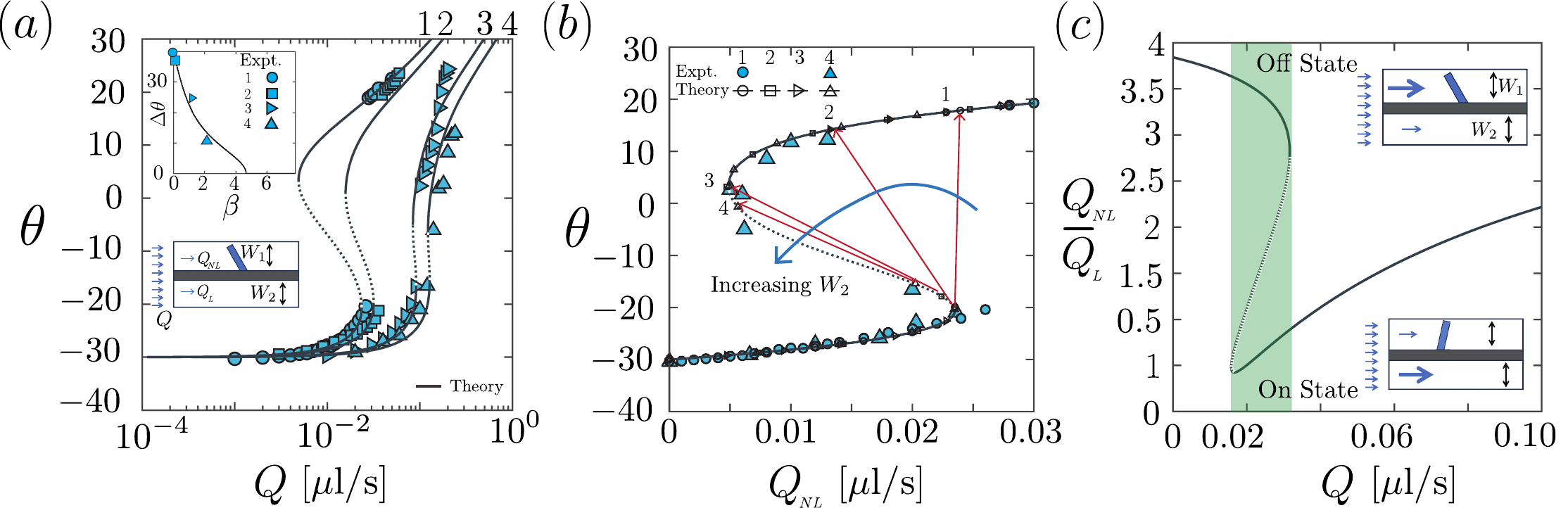} \caption{\textbf{A geometric load line tunes and suppresses the fluidic hysteron.} (a)~Flow--deformation curves for increasing bypass width ratio $\beta=W_2/W_1$, for cases 1--4 with $\beta=0$, $0.17$, $1.45$, and $3.13$, respectively. Increasing $\beta$ shifts snapping to larger imposed flow rates and reduces the jump amplitude $\Delta\theta$. Inset: $\Delta\theta$ decreases monotonically with $\beta$ and vanishes beyond a critical bypass width. (b)~Constitutive response plotted against the flow through the fiber channel, $Q_{NL}$. Solid lines show the theoretical prediction, illustrating how the bypass load line allows the system to traverse the negative-differential branch that is inaccessible in the single-channel limit. Symbols are particle-tracking-velocimetry measurements confirming the reduction of flow through the fiber channel with increasing deformation. (c)~Passive flow routing for case~2. Snapping switches the preferred flow path from the fiber channel to the bypass, so that the fiber channel acts as a relatively open conduit before snapping and a relatively closed one after.}
\label{fig:fig-3} 
\end{figure*}

\section{Geometric load-line control}

To turn the single fluidic hysteron into a tunable circuit element, we add a linear bypass channel of width $W_2$ in parallel with the fiber channel of width $W_1$, and define the control parameter $\beta\equiv W_2/W_1$. The limit $\beta=0$ recovers the isolated hysteron (Fig.~\ref{fig:fig-3} (a)). As $\beta$ increases, an increasing fraction of the imposed flow is diverted through the bypass. Snapping is shifted to larger imposed flow rates, while both the jump amplitude $\Delta\theta$ and the hysteresis width $\Delta Q$ decrease; beyond a critical value of $\beta$, snapping is completely suppressed (Fig.~\ref{fig:fig-3} (a) inset; Supplementary Information Section~5). This behavior follows from the shared pressure drop across the two branches. The total flow satisfies $Q = Q_{NL}(\Delta P,\mathrm{state}) + Q_L(\Delta P)$, where $Q_{NL}$ is the flow through the fiber channel and depends on both the pressure drop $\Delta P$ and the instantaneous fiber state (snapped or unsnapped), while the bypass flow is linear $Q_L = \Delta P / R_b(\beta)$. The bypass therefore provides a tunable load line: its linear flow--pressure relation intersects the nonlinear constitutive response of the fiber channel and selects the operating point. Decreasing the bypass resistance flattens the load line, progressively reducing the jump amplitude; beyond a critical $\beta$ the load line intersects the constitutive curve only once, eliminating bistability entirely (Fig.~\ref{fig:fig-3} (b)).

Functionally, the bypass converts the anchored fiber into a passive flow router. Before snapping, most of the flow passes through the fiber channel; after snapping, the large increase in its hydraulic resistance diverts the flow into the bypass (Fig.~\ref{fig:fig-3} (c)). The device therefore switches between relatively open and relatively closed flow states without requiring intrinsic elastic multistability, external actuation, or membrane-based valving \cite{gomez2017passive,garg2024passive}.

\section{Interacting fluidic hysterons and avalanches}

\begin{figure*}[tbh!]
 \centering
\includegraphics[width=1.0\linewidth]{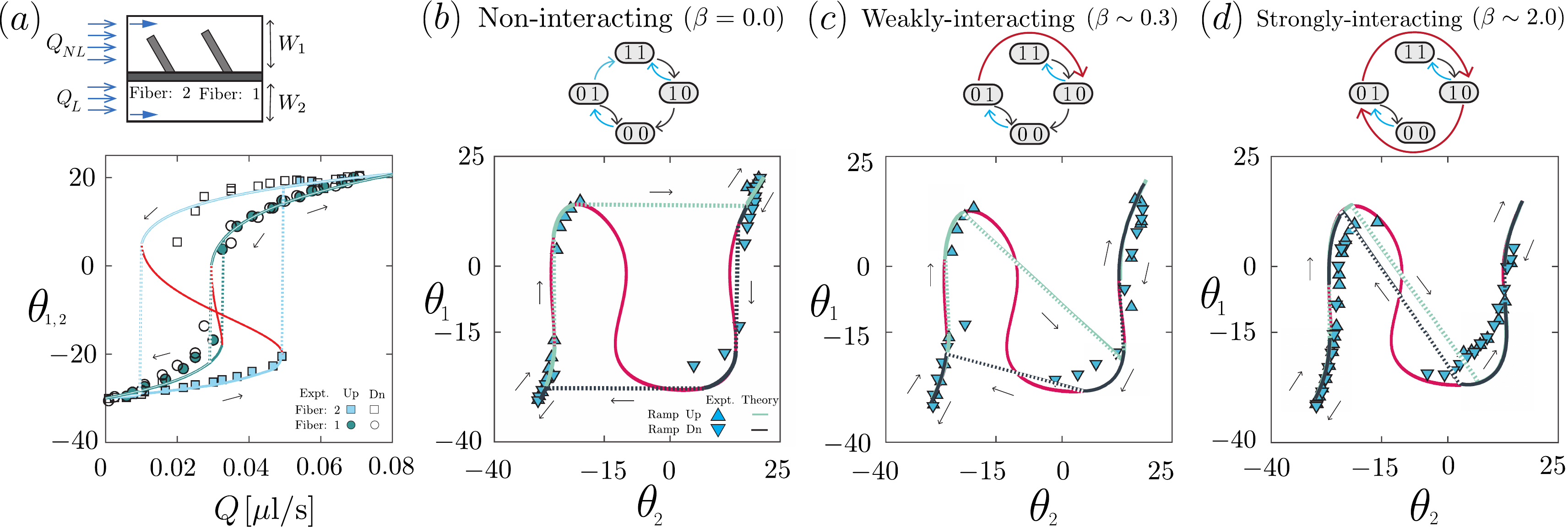}
\caption{\textbf{Transition pathways in a two-hysteron fluidic network.} (a)~Schematic of two fibers coupled through a parallel bypass channel with width ratio $\beta=W_2/W_1$. The individual constitutive curves are nested, with fiber~1 switching inside the hysteresis loop of fiber~2. (b)~Non-interacting limit ($\beta=0$). The transition graph follows the classical Preisach sequence $00\rightarrow01\rightarrow11\rightarrow10\rightarrow00$, and the $\theta_1$--$\theta_2$ phase space contains four discrete clusters connected only by axis-aligned transitions. (c)~Weakly interacting regime. A single avalanche appears during ramp-up, producing one diagonal jump in the phase space ($n=1$). (d)~Strongly interacting regime. Avalanches occur on both ramp-up and ramp-down, producing two diagonal jumps and reversing the unsnapping order relative to the snapping order ($n=2$). Lines represent the theoretical model and symbols represent experimental data.}
\label{fig:fig-4} 
\end{figure*}

Equipped with a tunable fluidic hysteron, we now move from single-element bistability to the collective switching of coupled hysterons \citep{keim2019memory}. In the no-bypass limit ($\beta=0$), the two fibers experience the same imposed flow but do not interact, and therefore behave as independent hysterons (Fig.~\ref{fig:fig-4} (a,b)). Labeling the unsnapped state by $0$ and the snapped state by $1$, two fibers with nested hysteresis loops (fiber~1 nested within fiber~2) traverse the sequence $00\rightarrow 01\rightarrow 11\rightarrow 10\rightarrow 00$ under a full ramp-up/ramp-down cycle, as shown by the transition graph in Fig.~\ref{fig:fig-4} (b) \citep{Terzi20}. The $\theta_1$--$\theta_2$ phase space reflects this independence: the system visits four well-separated clusters corresponding to the four binary states, and transitions occur along the coordinate axes, indicating that only one fiber switches at a time.

Introducing a bypass channel couples the fibers through global hydraulic constraints. The two fibers remain spatially separated, but the state of either fiber changes the effective resistance of the nonlinear branch and thereby the flow seen by the other. The interaction is long-ranged and effectively all-to-all within the fiber branch, with strength controlled by the geometric parameter $\beta=W_2/W_1$.

For small $\beta$, the coupling already produces qualitatively new behavior. When one fiber snaps, $00\rightarrow 01$, the resistance of the fiber branch increases, diverting flow into the bypass and reducing the flow through the remaining unsnapped fiber. The second fiber therefore snaps only at a larger imposed flow. When it eventually does so, the additional increase in branch resistance reroutes enough flow to unsnap the first fiber, producing an avalanche-like transition $01\rightarrow 10$; as the flow rate increases further, the first fiber snaps up again arriving at $11$ state (Fig.~\ref{fig:fig-4} (c)). In the $\theta_1$--$\theta_2$ phase space, this appears as a diagonal jump rather than a sequence of axis-aligned transitions. We denote by $n$ the number of such avalanches, with $n=0$ for the non-interacting limit and $n=1$ for this weakly coupled regime.

At larger $\beta$, corresponding to stronger coupling, a second avalanche appears on ramp-down. The unsnapping order is now reversed relative to the snapping order, $ 11\rightarrow 10\rightarrow 01\rightarrow 00$, and the phase space shows two diagonal jumps (Fig.~\ref{fig:fig-4} (d)). In hysteron language, this is an ``antiferromagnetic-like" interaction: the snapped state of one fiber favors the unsnapped state of the other, producing avalanche-mediated pathways that are forbidden in the non-interacting Preisach limit \cite{van2021profusion,liu2024controlled,lindeman2025generalizing}.

\section{Collective memory and return-point-memory violation}

\begin{figure*}[tbh!]
 \centering
\includegraphics[width=0.8\linewidth]{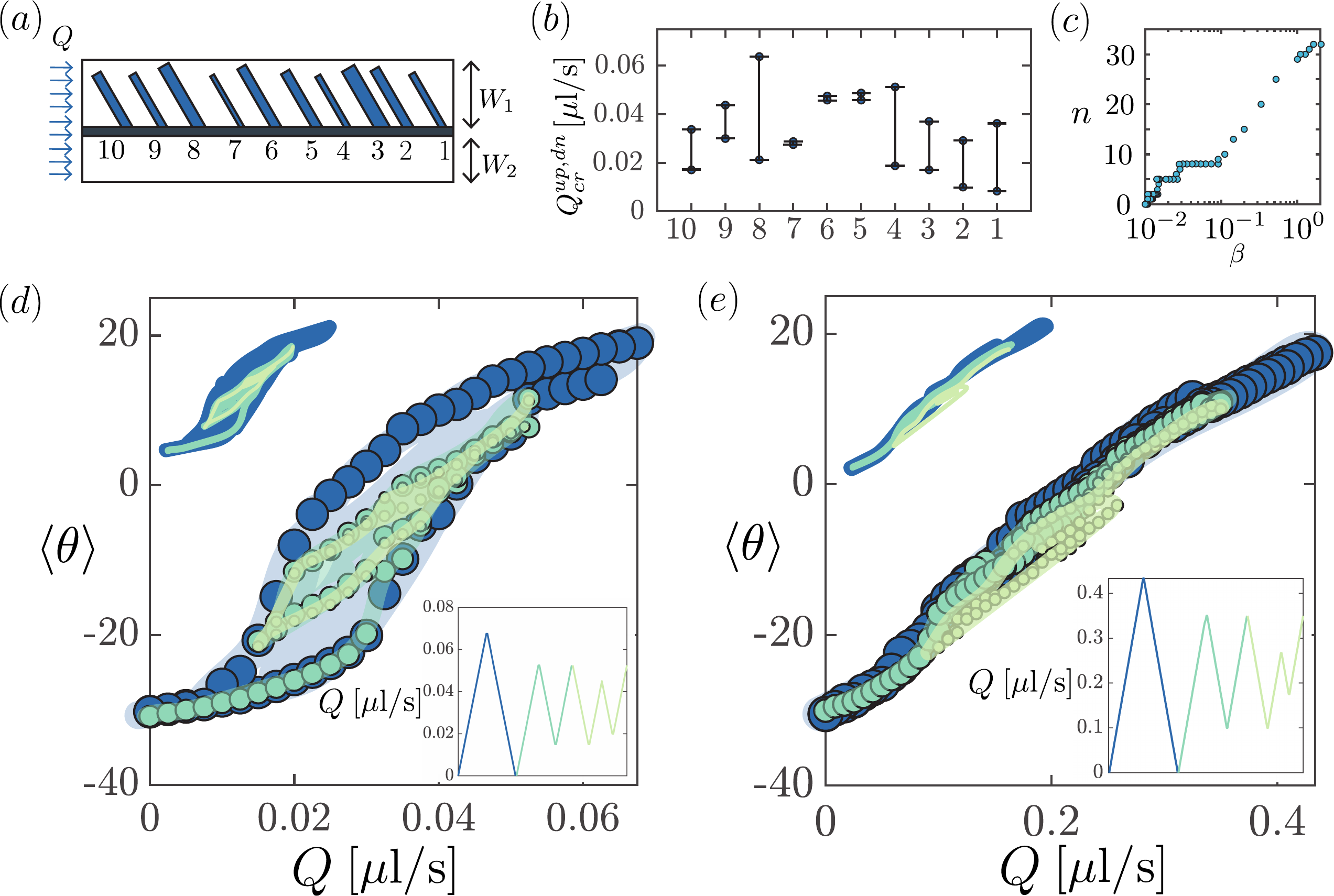}
\caption{\textbf{Collective memory in arrays of fluidic hysterons.} (a)~Schematic of an array of $N=10$ fibers connected in series with a parallel bypass channel. (b)~Measured snap-up and snap-down thresholds, $Q_{cr}^{up}$ and $Q_{cr}^{dn}$, for each of the ten fibers, showing the spread of individual switching flow rates. (c)~Number of avalanches $n$ as a function of the bypass width ratio $\beta$. Stronger interactions produce progressively more avalanches per cycle. (d)~Mean fiber angle $\langle\theta\rangle$ under nested write--read protocols in the non-interacting limit ($\beta=0$). Minor loops remain enclosed within the major loop, demonstrating return-point memory (RPM). Inset: quasistatic flow-driving protocol. (e)~Mean fiber angle $\langle\theta\rangle$ for an interacting network ($\beta\approx0.3$). Minor loops extend beyond the major loop and the system fails to return to previously visited states, demonstrating violation of RPM. Inset: quasistatic flow-driving protocol.}
\label{fig:fig-5} 
\end{figure*}

To probe collective memory, we study arrays of $N=10$ fibers in series under standard write--read protocols (Fig.~\ref{fig:fig-5} (a,b)), in which the drive is reversed at a sequence of nested turning points and then returned to earlier extrema to test whether the system recovers its prior state (Fig.~\ref{fig:fig-5} (d,e) inset). In the non-interacting limit ($\beta=0$), the array exhibits return-point memory (RPM): minor loops remain enclosed within the major loop, and returning to a previously visited extremum restores the corresponding earlier state \citep{keim2019memory}. This is observed both in the mean tip angle $\langle\theta\rangle$ and in the average binary state $\langle s\rangle=\frac{1}{N}\sum_i s_i$, where $s_i=0,1$ denotes the unsnapped or snapped state of the $i$th fiber (Fig.~\ref{fig:fig-5} (d) and Supplementary Information Section~6).

Introducing bypass-mediated interactions ($\beta>0$) destroys RPM. Avalanches allow the system to access transition pathways that are forbidden in the non-interacting limit (Fig.~\ref{fig:fig-5} (c)), and minor loops extend beyond the major loop (Fig.~\ref{fig:fig-5} (e)): the system visits states not encountered on the initial ramp and fails to recover previously visited intermediate states upon reversal. This is consistent with the general expectation that interacting hysterons admit a profusion of transition graphs, many of which violate RPM; frustrated, antiferromagnetic-like interactions of the kind produced by our bypass coupling provide a particularly direct route to such violations \cite{van2021profusion,shohat2025geometric,lindeman2025generalizing}. The bypass ratio $\beta$ therefore provides a single geometric control parameter that continuously tunes the network from Preisach to non-Preisach collective behavior.

\section{Discussion}

Together, these results establish a fluidic realization of the hysteron framework developed in solid-state mechanical systems \cite{keim2019memory,Paulsen25}. Two recent studies have explored related territory in network analogues and fluidic networks: Altman et al.~\citep{Altman26} realized collective memory states and avalanches in an electronic analogue of bistable flow networks while Martinez-Calvo et al.~\citep{martinez2024fluidic} developed fluidic memristors, where hysteresis emerges collectively from volume storage and nonlinear resistance. By contrast, in our system a single fiber already acts as a hysteron. Its bistability is geometric in origin: the snapping transition has the structure of a cusp catastrophe, controlled by the length ratio $\lambda$ and anchor angle $\theta_0$, while stiffness sets the flow scale and tunes the width of the hysteresis window. Cusp catastrophes are generic organizing structures for the onset of bistability in nonlinear systems, from mean-field first-order transitions to optical bistability \cite{Poston78,Lugiato84}; here, this structure emerges from the coupling between viscous flow and elastic deformation rather than from an intrinsic elastic energy landscape.

The fluidic setting also changes the nature of interactions between hysterons. In solid-state systems, coupling is typically mediated by short-ranged elastic stresses or contact forces. Here, a local change in resistance redistributes pressure and flux throughout the network through Kirchhoff-like hydraulic constraints. In the presence of a bypass, this leads to an all-to-all coupling within the fiber branch, with interaction strength controlled by the single geometric parameter $\beta=W_2/W_1$. The resulting antiferromagnetic-like interaction, in which snapping of one fiber disfavors snapping of the others, generates avalanches and violates return-point memory, consistent with the predicted proliferation of transition pathways in frustrated hysteron networks \cite{van2021profusion,muhaxheri2024bifurcations,liu2024controlled,lindeman2025generalizing}. Our fluidic setting therefore provides a tunable, optically resolved realization of interacting-hysteron physics, in which interaction strength is set by geometry rather than by local contact mechanics or material disorder.

The ability to tune both the local constitutive response and the interaction strength using geometry suggests a broader design principle for memory-bearing flow networks. Embedding fluidic hysterons in networks beyond a simple linear series, such as loops, trees, or disordered graphs, would introduce topology-dependent interactions and provide experimental access to questions about how network architecture shapes transition pathways, frustration, and memory capacity \citep{Bhattacharyya22,Bhattacharyya23,shohat2025geometric}. In this sense, fluidic hysteron networks provide a complementary experimental setting for the statistical physics of interacting hysterons, with interactions mediated by conservation laws rather than elastic contact or stress redistribution.

Finally, this system provides a minimal physical analogue of living flow networks, which couple local mechanical adaptation, such as compliance and shear-dependent remodeling, to global hydraulic constraints \citep{Serandour24}. Real biological networks are far richer, involving active remodeling, growth, and biochemical signalling. Yet our results show that passive ingredients alone suffice to generate history-dependent collective response, identifying flow--structure feedback as a sufficient mechanism for memory in fluidic networks. A central open question is therefore which aspects of memory and adaptation in living and soft adaptive flow networks genuinely require active control, and which are already latent in the passive mechanics of coupled deformable elements under global constraints.

\noindent\textit{Note added.}---After submission of this manuscript, Stinissen \textit{et al.} reported two intrinsically bistable inflatable membranes coupled by an equal-volume-change constraint, with preset volume selecting the pair's state-diagram topology and whether avalanches occur \cite{Stinissen26}. By contrast, the hysterons here are elastic fibers that are not intrinsically bistable: bistability arises from through-flow elastohydrodynamic feedback, and interactions from bypass-mediated hydraulic redistribution in a flow network.

\section{Methods}
\subsection{Device fabrication}
Experiments were performed in polydimethylsiloxane (PDMS) microfluidic devices fabricated by standard soft lithography. Channel moulds were patterned on silicon wafers using a laser writer. A PDMS prepolymer and curing agent mixture (10:1 by weight) was poured onto the mould and cured for 4~h at $65\,^\circ\mathrm{C}$. The cured PDMS slab was then bonded to a glass slide spin-coated with a thin layer of uncured PDMS mixture. The channels used in this study had widths $W=800,\,1200,\,1600,$ and $2000~\mu\mathrm{m}$ and a height $h_c=80~\mu\mathrm{m}$ (Fig.~\ref{fig:fig-1} (c)).

Deformable hydrogel fibers were fabricated in situ using stop-flow lithography \cite{dendukuri2007stop}. One wall of each channel contained PDMS microposts that served as anchors for the fibers. A photosensitive solution of PEGDA, Milli-Q water, and photoinitiator was injected into the channel using a syringe pump. The flow was then stopped and the solution exposed to masked ultraviolet light, photopolymerizing the hydrogel structures. This procedure yielded fibers with Young's modulus $E=0.30\pm0.04~\mathrm{MPa}$, height $h=70~\mu\mathrm{m}$, width $b\in[100,400]~\mu\mathrm{m}$, and length ratio $\lambda=L/W\in[0.65,1.00]$. Fibers were anchored at upstream tilt angles $\theta_0\in\{-50^\circ,-40^\circ,-30^\circ,-20^\circ\}$. Some devices additionally contained a parallel bypass channel of width $W_2$ alongside the fiber channel of width $W_1$, characterized by the width ratio $\beta=W_2/W_1$ (inset of Fig.~\ref{fig:fig-3} (a)).\\

\subsection{Flow control and imaging}
The volumetric flow rate $Q$ was imposed using a Harvard Apparatus syringe pump. Quasistatic ramps in the range $Q\in[10^{-3},10^{0}]~\mu\mathrm{L\,s^{-1}}$ were chosen to ensure negligible inertia ($Re\ll1$). Fiber deformation was imaged in top view at 1~Hz using brightfield microscopy. The tip angle $\theta$ was extracted from binary masks of the fiber centerline using custom image-analysis code based on intensity thresholding. Uncertainty was estimated from repeated segmentation and calibration.\\

\subsection{Reduced elastohydrodynamic model}
The fiber was modeled as an inextensible Euler--Bernoulli beam with linear constitutive law $M=EI\kappa$, clamped at the anchor and free at the tip. The surrounding flow was decomposed into parallel pathways: a variable-height tip-gap pathway with hydraulic resistance $R_g(\theta)$ and constant-thickness lubrication-layer pathways with resistance $R_i$. Each pathway was described using lubrication approximations appropriate for slender gaps (Supplementary Information Section~2). Flow partitioning between these pathways determined the pressure and shear tractions acting on the fiber, which were projected onto an effective distributed load in the beam equation. Steady states were obtained by solving the coupled nonlinear elastohydrodynamic system for the tip angle $\theta$, and, when needed, the full fiber shape. Stability was assessed by identifying segments of the constitutive curve with $dQ/d\theta<0$.\\

\subsection{Elastoviscous number and geometric parameters}
The length ratio is $\lambda=L/W$. A geometric prefactor $\Lambda=\frac{3b(h_c+h)}{h_i^2}$ collects the remaining geometric parameters. The elastoviscous number is defined using the mean velocity in the lubrication layers, $U_m$, as $\mathrm{El}=\frac{\Lambda\mu U_mL^3}{D}$, where $D=EI$ is the flexural rigidity. For long fibers, $U_m$ was inferred from the hydraulic partition model using the measured flow rate $Q$ and device geometry. A corresponding characteristic flow scale, $\tilde{Q}:=\frac{6q_i}{\mathrm{El}} =\frac{3Dh_i}{\Lambda\mu L^2}$, combines the elastic and viscous contributions into a single control scale used to nondimensionalize the hysteresis width in 
Fig.~\ref{fig:fig-2} (c) (Supplementary Information Section~2).\\

\subsection{Transition graphs, state labels, and RPM protocols}
Each fiber was assigned a binary state $s_i\in\{0,1\}$ according to whether $\theta_i$ lay on the lower (unsnapped) or upper (snapped) branch. The switching thresholds $Q_{cr}^{up}$ and $Q_{cr}^{dn}$ were determined from ramp-up and ramp-down experiments. Transition graphs were constructed by listing the stable states and directed transitions under monotonic increases or decreases of $Q$. The return-point-memory protocols in Fig.~\ref{fig:fig-5} (d,e) employed nested turning points $Q_1<Q_2<\dots$, with controlled partial cycles; the readout compared the recovered state at repeated extrema. The collective state of an array was characterized by the mean angle $\langle\theta\rangle:=\frac{1}{N}\sum_i \theta_i$ and the mean binary state $\langle s\rangle:=\frac{1}{N}\sum_i s_i$, where $\theta_i$ and $s_i$ are the tip angle and binary state of the $i$th fiber, respectively.\\

\section*{Data availability}
All data and analysis codes are available upon publication.

\bibliography{Reference}
\bibliographystyle{unsrt}
\end{document}